\documentclass[pdflatex,sn-vancouver,Numbered]{sn-jnl}

\usepackage{graphicx}%
\usepackage{multirow}%
\usepackage{amsmath,amssymb,amsfonts}%
\usepackage{amsthm}%
\usepackage{mathrsfs}%
\usepackage[title]{appendix}%
\usepackage{xcolor}%
\usepackage{textcomp}%
\usepackage{manyfoot}%
\usepackage{booktabs}%
\usepackage{algorithm}%
\usepackage{algorithmicx}%
\usepackage{listings}%
\usepackage{longtable}%
\usepackage{gensymb}
\usepackage{comment}
\usepackage{enumitem}
\usepackage{hyperref}

\begin{document}

\title[Article Title]{
Selection of CMIP6 Models for Regional Precipitation Projection and Climate Change Assessment in the Jhelum and Chenab River Basins
}

\author*[1,2]{\fnm{Saad Ahmed} \sur{Jamal}}\email{saad.jamal@uevora.pt
}

\author[3]{\fnm{Ammara} \sur{Nusrat}}

\author[4]{\fnm{Muhammad} \sur{Azmat}}

\author[5]{\fnm{Muhammad Osama} \sur{Nusrat}}



\affil*[1]{\orgdiv{Department of Computer Science, University of Évora, 7000-671 Évora, Portugal}, \country{Portugal}}
\affil*[2]{\orgdiv{MED—Mediterranean Institute for Agriculture, Environment and Development \& CHANGE—Global Change and Sustainability Institute, Institute for Advanced Studies and Research, University of Évora, 7004-516 Évora}, \country{Portugal}}

\affil[3]{\orgdiv{NUST Institute of Civil Engineering (NICE), School of Civil and Environmental Engineering (SCEE), National University of Sciences and Technology (NUST), Islamabad}, \country{Pakistan}}

\affil[4]{\orgdiv{Institute of Geographical Information Systems (IGIS), School of Civil and Environmental Engineering (SCEE), National University of Sciences and Technology (NUST), Islamabad}, \country{Pakistan}}

\affil[5]{IMT Atlantique, Lab-STICC UMR CNRS 6285, Brest, France}




\abstract{
Effective water resource management depends on accurate projections of flows in water channels. 
For projected climate data, 
use of different General Circulation Models (GCM) 
simulates contrasting 
results. 
This study shows selection of GCM for the latest generation CMIP6 
for hydroclimate change impact studies.
Envelope based method was used for the selection, which includes components based on machine learning techniques, allowing the selection of GCMs without the need for in-situ reference data. 
According to our knowledge, for the first time, 
such a comparison was performed for the CMIP6 Shared Socioeconomic Pathway (SSP) scenarios data. 
In addition, the effect of climate change under SSP scenarios was studied, along with the calculation of extreme indices. Finally, GCMs were compared to quantify spatiotemporal differences between CMIP5 and CMIP6 data.
Results provide NorESM2\_LM, FGOALS\_g3 as selected 
models for the Jhelum and Chenab River. 
Highly vulnerable regions under the effect of climate change were highlighted through spatial maps, which included parts of Punjab, Jammu, and Kashmir.
Upon comparison of CMIP5 and CMIP6, no discernible difference was found between the RCP and SSP scenarios' precipitation projections. In the future, more detailed statistical comparisons could further reinforce the proposition.
}




\maketitle

\section{Introduction}\label{sec1}


Climate Change is a major challenge to the world.
United Nations has announced 17 Sustainable Development Goals in which Climate Change is one of them. To ensure a sustainable world, all the countries of the world must cooperate to avoid environmental degradation. Global Warming is a fact that cannot be denied \citep{GlobalWarminghoughton2009global}. It is primarily driven by the rising concentration of specific greenhouse gases in the atmosphere. Carbon dioxide is the main constituent that is responsible for the maintenance of temperature over the globe. It constitutes 0.04\% of air only. A slight change in its concentration can lead to disruption of the ecosystem. 
Projections are made upon the concentration of gases in the atmosphere. 
These experiments produce different results. The aim of this study is to determine which results are more reliable by quantifying uncertainties.
Extreme rainfall events over the past few decades have resulted in severe damage in Pakistan. Regionalization can be done based upon extreme rainfall events \citep{qamar_azmat_rainfall_2017}. Such regionalization helps in a better understanding of the local climate.
\citet{https://doi.org/10.1002/2014WR015549} investigated the factors contributing to uncertainty in the forecasts of hydrological regimes. The main sources of uncertainty came out to be input Hydro-Climate Data and uncertainties due to hydrological modeling and discharge projections. \citet{mendlik_selecting_2016} Provided a method for selecting General Circulation Models through the use of Machine Learning that will be used in this study for the GCM selection. \citet{melsenhess-22-1775-2018} did a GIS-based study on mapping these uncertainties geographically. This further enhances the applicability of calculated uncertainties according to their location. In Europe, climate projection uncertainties are calculated through Ensembles or Coupled Model Intercomparison Projects. The researcher made use of explanatory methods to generate easy-to-comprehend uncertainty maps proposed by \citet{melsenhess-22-1775-2018}.

CMIP stands for Coupled Modeled Intercomparison Project launched by the World Climate Research Program (WRCP) \citep{stocker2014climateIPCCFifthGeneration}. Based on historical data, 16 countries have participated in its development for future scenarios. It provides simulated model output called General Circulatory Models which are created according to WRCP standards. This output contains modeled Atmosphere, Land, Snow, Ocean related variables for the globe. In this study, atmospheric variables, precipitation, and temperature will be used. CMIP6 data was published in 2020, while some of its GCMs are still under development. Online available sources of the latest generation of General Circulatory Models are under observation.
CMIP aims to develop global simulations of the coupled climate system and provide a diverse array of model outputs to enhance comprehension of historical, current, and future climate variability.
While CMIP5 models show enhancements in process representation compared to CMIP3, significant uncertainty persists regarding future climate, and this uncertainty may even escalate locally due to the increased number of available models \citep{Lutz_Immerzeel_https://doi.org/10.1002/joc.4608, joetzjer_present-day_2013}.
Historical data of CMIP6 based upon extreme precipitation was studied by \citet{SRIVASTAVA2020100268}. In this study, consequent research for projected CMIP6 data evaluation was proposed. Their accuracy, spatial resolution, spatial extent, and time period of data availability were analyzed to minimize the volume of data that needs to be downloaded. Such signals are particularly useful in future natural disaster-related studies. 
\citet{Cookhttps://doi.org/10.1029/2019EF001461}] developed a methodology to study drought using CMIP6 data. Additionally, we examined sources of errors in models and available methods for calculation. This study implements an effective way to demonstrate these uncertainties.
Python code was developed for automated downloading for each scenario. The data format available on the server was (.nc). Server-based sub-setting was applied for the known coordinates of the study such that it covers the whole study area and also reduces the downloading of unnecessary data. The Python code will be shared for scientific purposes upon reasonable request. 

The NEX-GDDP-CMIP6 dataset originates from downscaled projections based on General Circulation Models (GCMs), as part of Phase 6 of the Coupled Model Intercomparison Project (CMIP6). These predictions are created using various scenarios known as Shared Socioeconomic Pathways (SSPs). The GCMs were designed to support the Sixth Assessment Report (AR6) of the Intergovernmental Panel on Climate Change (IPCC). The dataset offers high-resolution, bias-corrected climate forecasts on a global scale, essential for understanding the impact of climate change on regional processes like topography and climate. This data, accessible through the 
NASA Center for Climate Simulation (NCCS) portal, 
provides daily records to evaluate the potential effects of climate change across different environmental parameters. CMIP6 showcases advancements over its predecessors in climate model intercomparison \citep{CMIP6isBetter_https://doi.org/10.1029/2022EF002972}.


Agriculture serves as the backbone of Pakistan's national economy as it contributes 34\% of the total Gross Domestic Product (GDP). Agriculture-based products fetch 80\% of the country’s total export earnings, and according to the Agriculture Department of the Government of Punjab, more than 48\% of the labor force is engaged in this sector \citep{amjad2008does}. 
Water resources of 
the region are depleting rapidly \citep{AZMAT2018961}. The per capita water availability in Pakistan has seen a significant decline, dropping from nearly 5,260 cubic meters in 1951 to around 1,040 cubic meters by 2010, even as the population reached approximately 190 million during that period \citep{Latif_Haider_Rashid_2016}. This situation will further worsen with the projected increase in population of about 230 million by 2025, with only 800 cubic meters of water left. That means that by 2025, our available water resources will not be able to meet our needs. Agriculture alone is drawing 80\% of the total consumed water and our irrigation efficiency is less than 30\% \citep{ul2021critiqueAgricultureWaterManagement}. 

Pakistan has been facing heavy flooding for many years. The impact of global warming and climate change has amplified the severity of floods. The 2010 flood in Pakistan serves as a prime example, as its likelihood of occurrence would have been much lower if climate change had not been a contributing factor. With the effect of Climate Change, such events have a greater possibility of occurrence in the future. Planning to cope with such events needs to be done before the occurrence of such disasters. Lack of Infrastructure and a weak economy further add to the 
need to research such a topic.  
Therefore, a detailed analysis of CMIP data could provide potentially suitable models that can be used in future studies on the selected rivers. The output of this research will support the development of the Flood Management System of Pakistan and National Spatial Data Infrastructure (NSDI). Projected river discharges can be used to assess the balance between water supply and demand for irrigation as well as for hydropower production.

\begin{figure}[H]
    \centering
    \includegraphics[width=8cm]{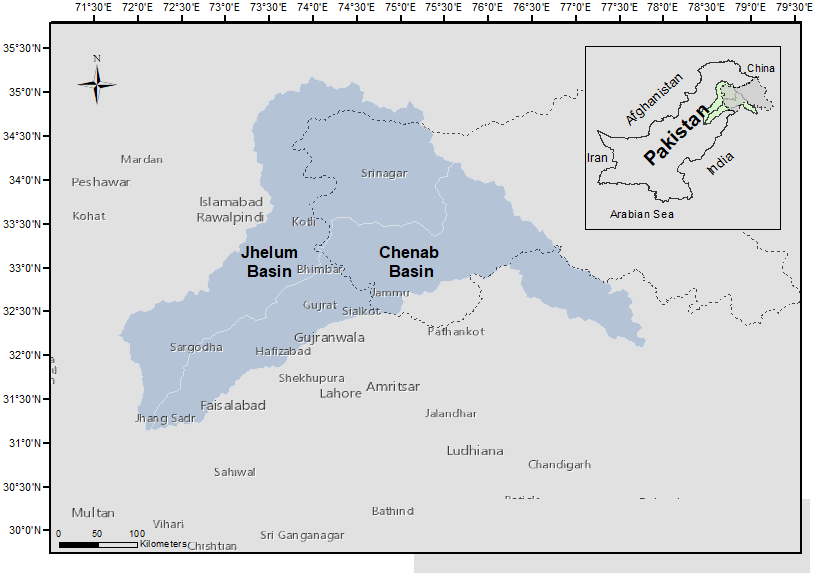}
    \caption{Map of Jhelum Chenab River Basin    }
    \label{fig:MapStudyArea}
\end{figure}

The study area was taken as Jhelum and Chenab Rivers, which are transboundary rivers in the subcontinent.
The Jhelum and Chenab River basin was selected as the study area due to frequent flooding and the lack of extensive research already done in the basin. Recently, CMIP6 data has been made available for the region, which can be used for flood and hydrological research. 
Also, the selection was made because the authors had ground knowledge about these river basins. Figure 
\ref{fig:MapStudyArea} shows watershed delineation (catchment basin) for the Jhelum and Chenab Rivers.
Watershed delineation for the study area was performed in ArcGIS using the Arc Hydro Toolset, which requires selecting an outlet point on the river. The outlet was set near the Trimmu barrage since it is located at the confluence point of the two rivers, so to delineate the complete river catchment, the point was digitized. First, the procedure was carried out for the Jhelum basin, and then the same procedure was applied to the Chenab catchment. SRTM digital elevation model elevation was used to extract elevation for the data extraction points within the study area. Both the river basins were merged in order to have a single basin for further use, both in the upcoming hydrological modeling.


The use of GIS with remote sensing data is essential for producing reliable results in climate change studies.
Due to industrialization and the overuse of fossil fuels, the concentration of GHGs has been increasing at an alarming rate.  
This study is essential for generating future climate projections and implementing preventive measures to ensure a safer world, regardless of whether emissions remain constant, increase, or decrease.
Several previous studies in the selected area used GCMs without scrutinizing them for regional relevance, directly importing the data into hydrological modeling processes. As a result, significant uncertainties remain unaddressed. 
This assumption within model-based studies can be sorted out through quantification of error terms or the selection of GCMs for the study area. Using Geoinformatics tools, it is possible to come up with adequate results by applying modern techniques. When studies related to projected floods and water-related hazards are being done, General Circulation Models uncertainty remains skeptical. Through the use of knowledge of spatial analysis and programming skills, one can come up with ways to solve such a problem. Particularly, with the use of high-level programming languages such as Python or R language and Geographical Information Systems (GIS), handling of big atmospheric data is possible. 
When information regarding the selection of CMIP6 GCMs and their applicability to the Jhelum and Chenab Rivers is available, hydrologists can make more reliable future projections. Using this framework, a similar methodology can be applied to other rivers in Pakistan, such as the Indus, Kabul, or Ravi, to conduct optimal studies related to GCMs.
Uncertainty quantification can significantly help hydroclimatic researchers produce accurate statistics and rational results by comparing state-of-the-art hydrological models and artificial neural networks for flood analysis. 
After a thorough literature review, the research gap was explored. 

This study explores the following questions:
Which CMIP6 model is most appropriate for the Jhelum and Chenab River Basins?
Are the selected GCMs selected through extreme indices similar to ones selected through an envelop-based approach?
What is the extent of variability of CMIP6 models 
from CMIP5 models? 
Through literature, it was known that the selection of the CMIP6 model has not been done for the study area yet, nor does there exist a comparison between CMIP6 and CMIP5 data. 
Therefore, a methodology was formalised for this research gap addressing the following objectives. 

\begin{enumerate}[label=(\alph*)]
\item Computation of extreme precipitation indices for the latest generation CMIP6 data.
\item 
Selection of CMIP6 models producing extreme climate signals for Jhelum and Chenab River Basin.
\item 
Analyzing the variability of Shared Socioeconomic Pathways (SSP) and Representative Concentration Pathways (RCP) scenarios.

\end{enumerate}

This research leverages the high-resolution CMIP6 dataset, marking its first application in predicting extreme precipitation indices for the Jehlum and Chenab basins. It analyzes historical data spanning from 1950 to 2015 and explores two future scenarios based on Shared Socio-Economic Pathways (SSP245 and SSP585), with projections extending from 2015 to 2100. 
The results serve as a base for further research related to trend detection and climate modeling. Also, the results have applications in Watershed Management, Flood Hazard and Disaster Mitigation, Frequency Analysis, Big Atmospheric Data Handling, and Drought Monitoring.

\section{Literature Review}
\label{sec:theoretical_background}

A comparison of CMIP6 and CMIP5 data performance for the North Pacific region was made by \citet{CHEN2021100303}. Similarly, \citet{Xinhttps://doi.org/10.1002/joc.6590} provided such a comparison of monsoon precipitations for the East Asian region. The CMIP5 GCM data was downscaled and bias corrected for two projected scenarios, known as Representative Concentration Pathways (RCP4.5 and RCP8.5). A similar procedure for CMIP6 Projected Scenarios was also applied. But in addition, it has SSP Scenarios known as Social Economic Pathways. Data for SSP245, and SSP585 represent the current and extreme cases respectively that the world can go through in coming decades. \citet{SRIVASTAVA2020100268} performed statistical downscaling of GCM’s for South Asia that 
can be used as input data for the hydrological modeling. 

IFAS Model Data requirements have been studied, and data acquisition has been made to process the initial model requirement, e.g., the Digital Elevation Model (DEM). It is a data-extensive model. The available ground truth data of measurement stations in the region 
is not sufficient to run this model. An alternative approach for future climate impact modeling for the scarcely gauged basin as described by  \citet{AZMAT2018961} has better applicability within the study area.
Floods in the Indus river under RCP projections were also calculated \citep{WinjngaardRene10.1371/journal.pone.0190224}. 

\citet{khan_azmat_development_2023} did quantile mapping and linear scaling of satellite-based data with observed data to produce a bias-corrected dataset for the Pothohar plateau basin. 
\citet{jamal2023data} used data augmentation and data fusion to improve the accuracy of the machine learning model above the baseline.
Therefore, additional data has several applications that can be of benefit in various ways within the modeling process.

\citet{ali_performance_2023} findings revealed that NorESM2-MM, EC-Earth3-Veg, and MRI-ESM2-0 demonstrated exceptional accuracy among GCMs in replicating past precipitation extremes and forecasting future patterns.
The evaluation of General Circulation Models (GCMs) for accurately representing the annual spatial distribution of precipitation and temperature extremes (both maximum and minimum) was performed using the Kling-Gupta efficiency (KGE) metric \citep{salehie_selection_2023}. A comprehensive ranking of the GCMs was conducted by \citet{salehie_selection_2023} using a multi-criteria decision-making (MCDM) method to integrate the KGE values. The findings identified AWI-CM-1-1-MR, CMCC-ESM2, INM-CM4-8, and MPI-ESM1-2-LR as the models that most accurately simulated the observed climatic variables in the Amu Darya River Basin \citep{salehie_selection_2023}. Future projections from these models indicated an increase in precipitation by 9.9. 

\citet{IQBAL2021105525} conducted an evaluation of 35 Global Climate Models (GCMs) from the Coupled Model Intercomparison Project Phase 6 (CMIP6) to examine their ability to simulate rainfall patterns in Mainland South-East Asia (MSEA) between 1975 and 2014. The study employed Compromised Programming (CP) alongside four spatial statistical metrics to rank the GCMs. To pinpoint the models best suited to MSEA’s climatic characteristics, the Jenk natural break classification method was applied. This classification approach was similarly utilized by \citet{ahmad_impact_2021} in water quality impact assessments.

The study found that most CMIP6 GCMs effectively represented rainfall patterns in MSEA, though their performance varied across different statistical criteria. By using the compromised programming method to integrate multiple metrics, the study identified MRI-ESM2-0, EC-Earth3, and EC-Earth3-Veg as the most appropriate models for predicting rainfall in the region. These models were able to replicate the annual mean rainfall patterns in central and southern MSEA with a bias of less than 25\%. However, higher biases were observed in Myanmar’s western coastal region, where rainfall is highest, with values ranging from -25\% to -75\%. Once these biases are corrected, the selected CMIP6 models can be reliably used for climate projections and impact studies in MSEA. The primary aim of \citet{IQBAL2021105525}'s research was to rank the GCMs based on their closeness to Aphrodite data in the region.

\citet{cli6040089} performed the selection and downscaling of Representative Concentration Pathway (RCP) scenarios for the Upper Indus Basin. In the current research, selection was performed for Shared Social Pathway (SSP) scenarios for Jhelum and Chenab, which fall in the Lower Indus Basin. The selection was performed based on three mean criteria, extreme signals and correlation check. In addition, mathematical models were used to improve the resolution of GCMs for the study area.
\citet{rasmus} showed that using principle components instead of raw climate data for statistical downscaling results in higher correlation values and lower root-mean-squared errors. 
The selection process relies on either considering the complete spectrum of changes in climatic variables projected by all available climate models or assessing the skill of climate models in simulating past climate \citep{Lutz_Immerzeel_https://doi.org/10.1002/joc.4608}.
The first approach involves including a comprehensive range of climate signals to choose GCMs, forming the foundation for the envelope-based selection method.
Our study also incorporates the limitation of previous studies, which averaged out the precipitation values over the entire study area for GCM selection.  
Furthermore, the results from our research can be used as input in applications such as accelerated flow modeling \citep{Prince_mendlik_selecting_2016} of streams to more advanced urban anomalies detection that use unsupervised classification techniques \citep{SadoonHAMMAD2023100848}.

Climate models demonstrate a close match with real-world observations, as noted by \citet{cli12060078}. 
For this topic, \citet{cli9090139} reviewed existing research on extreme weather events and how global warming is impacting them on the Iberian Peninsula. The authors predicted an increase in heavy precipitation by 7\% to 15\%, with a moderate rise in extreme precipitation of about 5\% by mid-century, followed by a decline to near zero by 2100. Additionally, they forecast a considerable decrease in wet days, ranging from 40\% to 60\%, with a noticeable trend toward drier conditions as the century progresses.

\citet{atmos14101497} applied eight commonly used indices to evaluate the performance of models in simulating extreme precipitation. Their results showed that most models effectively captured the observed trends. The GCM UKESM1-0-LL emerged as the most accurate model. All models displayed high accuracy in reproducing climatological means, particularly for total precipitation. Overall, the models UKESM1-0-LL, CESM2, MIROC5, MRI-ESM2-0, CMCC-CM2-SR5, and MPI-ESM-2-LR showed strong performance in replicating both the patterns and broader climatic characteristics of extreme precipitation.

NFDI4Earth, National Research Data Infrastructure for Earth System Sciences, is a consortium that caters to the digital requirements of Earth System Sciences, where scientists collaborate globally and across disciplines to comprehensively comprehend the operations and interconnections within the Earth system \citep{Bernard_Henzen_Degbelo_Nüst_Seegert_2023}. The primary goal is to tackle the myriad challenges associated with global change. NFDI4Earth operates through a community-driven approach, offering researchers access to Fair, Accessible, Interoperable, and Reusable (FAIR) Earth System data. Additionally, it provides a cohesive and openly accessible platform for innovative research data management and data science methodologies.

The creation of the NFDI4Earth consortium resulted from a detailed bottom-up process, incorporating plenary discussions, voluntary teams, a steering committee, and a coordination group
\citep{Bernard_Henzen_Degbelo_Nüst_Seegert_2023}.
Climatechart, a framework for visualization of precipitation and temperature time series for global data, was created by \citet{climatechart.netdoi:10.1080/17538947.2020.1829112}. 
The capabilities of this framework were explored. 
Using the climatechart.net web application, a chart was obtained for a point within the study area. 
The tool offers various climate analysis charts utilizing a standardized visualization concept, specifically designed for comparing diverse local climates.

\begin{figure}
    \centering
    \includegraphics[width=1\linewidth]{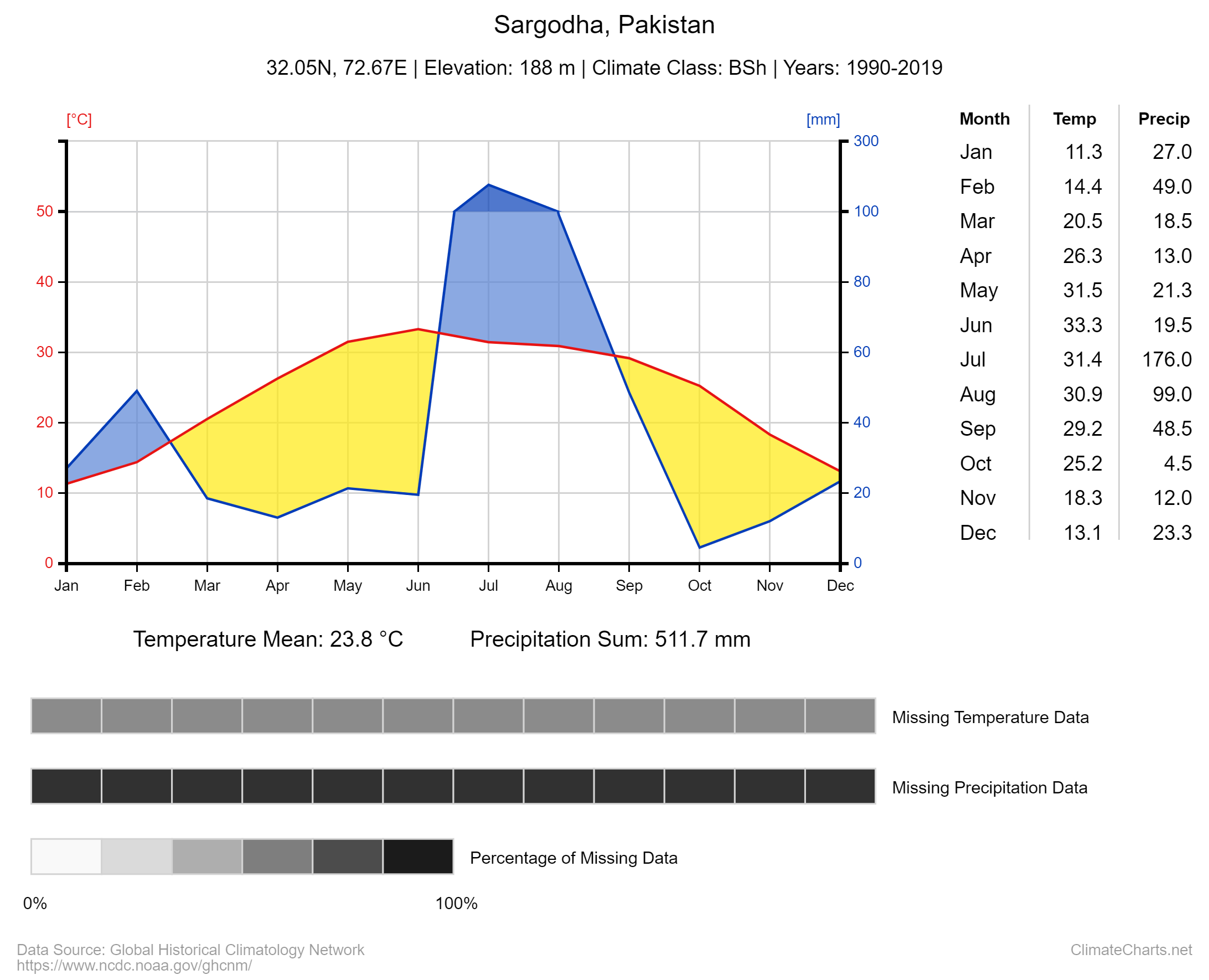}
    \caption{Climate chart for Sargodha, a point within the study area}
    \label{fig:ClimatechartSargodhA}
\end{figure}

Climatecharts.net utilizes two types of data: interpolated and simulated climate data for map grids, and actual weather information from global weather stations. It predominantly uses the Climatic Research Unit (CRU) Time-Series (TS) version 4.05, which provides high-resolution, gridded monthly climate data (January 1901 to December 2020), sourced from the NERC EDS Centre for Environmental Data Analysis. 
The dataset has a spatial resolution of 0.5$^{\circ}$  x 0.5$^{\circ}$  and temporal coverage from 1900 to 2020. The application empowers users to create personalized historical climate charts. Figure \ref{fig:ClimatechartSargodhA} shows climate details retrieved Climatechart.net for Sargodha, a point within the Study Area.

For higher resolution rainfall, Climate Hazards Group InfraRed Precipitation with Station (CHIRPS) data, climate extreme indices along with spatio-temporal trend analysis was carried by \citet{cli6040087Ghana} for a basin in Ghana.
The well-performing GCM serves as a good reference for further studies such as variability of droughts \cite{cli7060082Droughts}, flood, precipitation projections, trends \cite{cli11020033Trends}, climate risk and vulnerability assessment \cite{cli11110222RiskAndVulnerability}.
Extreme Climate Events based on such indices in the Himalayan Region were carried out by using the third generation CMIP3 models \citep{cli4010009ExtremesNorwestHimalaya}.

\citet{Nusrat_app10196878} performed the selection of CMIP5 GCMs for the Jehlum Chenab basin using the past performance method. The researchers also performed GCM selection using an envelope-based method \citep{atmos13020190}.  
The envelope-based technique is not unknown, it has been applied in various other processes as well. For example
PESA, an algorithm known as the Pareto Envelope-based Selection Algorithm, employs a straightforward hyper-grid-based approach to manage selection and diversity maintenance \citep{envlope10.1007/3-540-45356-3_82}.
In contrast to this study, NEX-GDDP-CMIP6 demonstrates improved proficiency in replicating the spatial patterns of extreme climate events, featuring higher spatial correlation coefficients and enhanced agreement among models
\citep{researchsquare1}.

In Pakistan, \citet{hess-23-4803-2019} evaluated CMIP5 GCMs using a spatial assessment matrix to rank their ability to simulate the spatial distribution of mean annual precipitation and seasonal rainfall patterns, including monsoon, winter, pre-monsoon, and post-monsoon, as well as maximum and minimum temperatures. The study identified NorESM1-M, MIROC5, BCC-CSM1-1, and ACCESS1-3 as the most proficient GCMs, while IPSL-CM5B-LR, CMCC-CM, and INMCM4 were categorized as having the weakest performance in this regard.

\section{Methodology}
\label{sec:research_approach}

The primary crucial phase within this framework entails the classification of regions according to the climatic features of the research zone. Regionalization pertains to the categorization of the area into groups with comparable climates and the identification of unique climate zones. This method requires extensive climate data records from multiple stations within the research zone. In this research, the regionalization process involved using the daily rainfall dataset from APHRODITE, which spans 35 years. This choice was made following a comparative assessment of its performance against other gridded datasets \citep{Nusrat_app10196878}. The bias correction was done in Python using looping linear scaling of precipitation data. The code will be shared upon request through a GitHub repository.

Regionalizing climate statistics finds applications in various fields such as agriculture, forecasting hydrological extremes, and managing basins. 
The methods that are commonly employed to demarcate zones include 
spatial convenience, subjective and objective partitioning \citep{w11030570}. 
These methods commonly utilize multivariate analysis techniques like Principal Component Analysis (PCA), 
multivariate analysis techniques, such as correlation analysis and clustering, were employed to 
locate areas characterized by homogeneous climate \citep{rasmus}.
The geographical convenience method relies on arbitrary administrative boundaries, which can be somewhat misleading. Subjective and objective partitioning methods group meteorological sites with similar climate statistics to define regions.

\begin{figure}[H]
    \centering
    \includegraphics[width=1\textwidth]{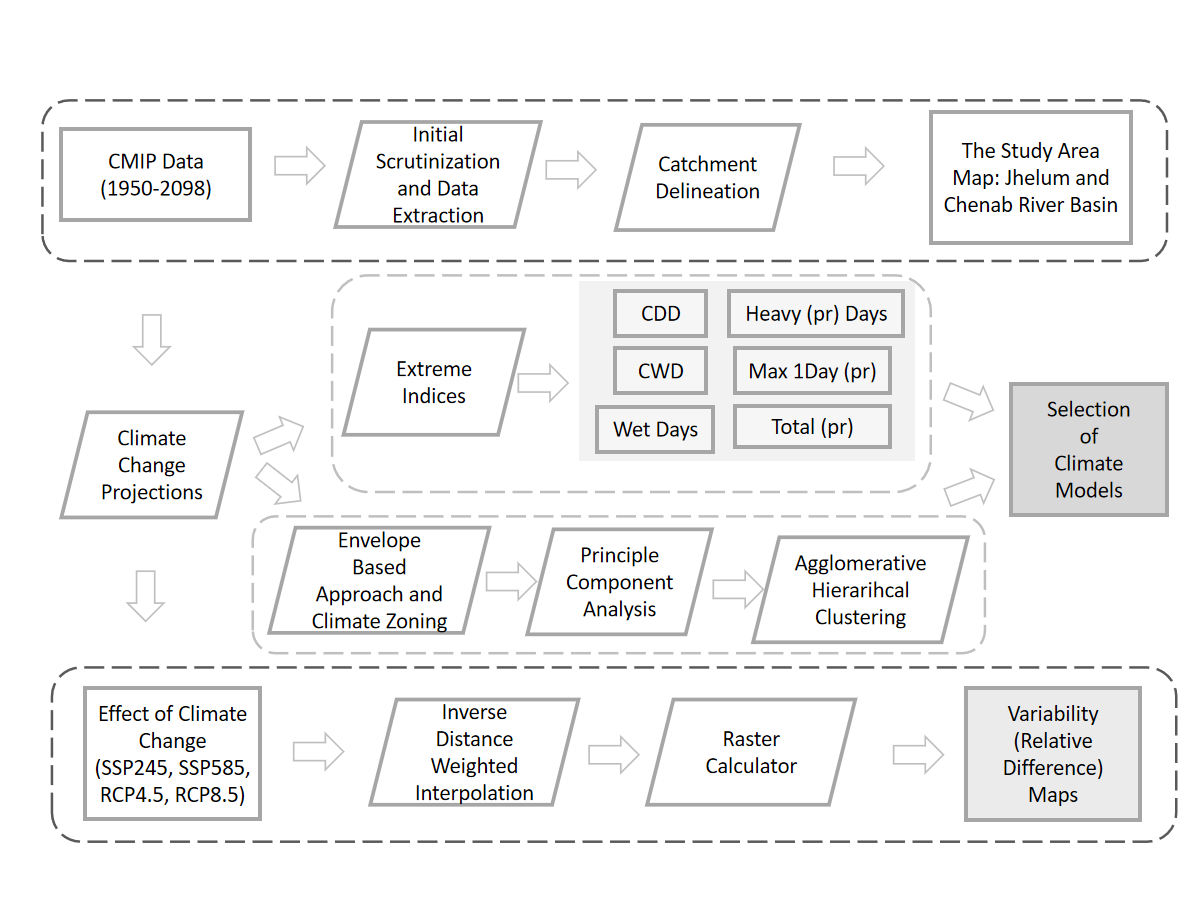}
    \caption{Methodological Flowchart}
    \label{fig:Methods1}
\end{figure}

In this research, regionalization was performed by applying agglomerative hierarchical clustering (AHC) to the principal components (PCs) extracted from precipitation data across nearby stations. Metrics for cluster validity were used to confirm the optimal number of clusters. The detailed process for regionalization is provided later in this section, following the explanation of the CMIP6 dataset used. Figure \ref{fig:Methods1} presents a sketch of the adopted methodology.

NCCS THREDDS Data Catalog was used to download the data. 
The latest generation of bias-corrected NEXGDDP CMIP6 data is available on the NASA Center for Climate Solutions website. The dataset was accessed using Python API. The Python code used for downloading data is made available online through the GitHub repository for this research.
The latest addition to accessing data includes the Amazon Web Services (AWS) Simple Storage Service (S3) bucket. However, for this research, the Python API option was used. 

The available data files were subsetted online by providing coordinates for the bounding box that covered the study area. Result, each file consisted of approximately 179 MB. 
For the historical scenario, the data downloaded spans from 1950 to 2014, which corresponds to 64 years of historical data.
For future scenarios, data downloaded spanned from 2015 to 2099. A total of 85 years of data for 23 GCMs was downloaded for SSP245 and SSP585. 
The downloaded data required 837 GB of disk space from the total available dataset size of 35.6 TB. Furthermore, the extracted data for 138 points of interest within this research further reduced the size to 2.78 GB. 
Out of 35 GCMs, 23 were scrutinized based on the completeness of data for a complete time span of the study. Table \ref{tab:GCMModelNames} presents the institution and resolution of GCMs.

\begin{longtable}{|p{4cm}|p{4cm}|p{2cm}|p{2.8cm}|}
    
    \caption{ 23 General Circulation Models from NEX-GDDP-CMIP6 chosen based on completeness, Source: \citep{CMIPGCMGROSE2023100368}
    }
        \label{tab:GCMModelNames}\\

    

\hline
\textbf{Model Name}	& \textbf{Institution/Country}	& \textbf{Horizontal Resolution (lon × lat)	 
}	& \textbf{Degree} \\
    \hline

ACCESS-CM2 	 & CSIRO-ARCCSS/Australia &	144 × 192	& 1.875$^{\circ}$ × 1.25$^{\circ}$  \\
ACCESS-ESM1-5  &	CSIRO/Australia	& 192 × 145	& 1.875$^{\circ}$  × 1.25$^{\circ}$  \\
BCC-CSM2-MR & BCC/China &	160 × 320	& 1.125$^{\circ}$  × 1.125$^{\circ}$  \\
CanESM5 	& CCCma/Canada	& 64 × 128	& 2.812$^{\circ}$  × 2.77$^{\circ}$  \\
CMCC-CM2-SR5 	&CMCC/Italy	& 288 × 192	& 1.25$^{\circ}$  × 0.9375$^{\circ}$  \\
EC-Earth3 	&EC-Earth-Consortium
&256 × 512	&0.703$^{\circ}$  × 0.703$^{\circ}$  \\
EC-Earth3-Veg-LR &	EC-Earth-Consortium
&	256 × 512	&1.125$^{\circ}$  × 1.125$^{\circ}$  \\
FGOALS-g3 &	CAS/China&	180 × 80&	2$^{\circ}$  × 2.025$^{\circ}$  \\
GFDL-CM4 	& NOAA-GFDL/USA	& 288 × 180	& 1.25$^{\circ}$  × 1$^{\circ}$  \\
GISS-E2-1-G &	NASA-GISS/USA&	144 × 90&	2.5$^{\circ}$  × 2$^{\circ}$  \\
INM-CM4-8 	&INM/Russia	&180 × 120	&2$^{\circ}$  × 1.5$^{\circ}$  \\
INM-CM5-0	&INM/Russia	&180 × 120	&2$^{\circ}$  × 1.5$^{\circ}$  \\
IPSL-CM6A-LR 	&IPSL/France	&144 × 143	&2.5$^{\circ}$  × 1.259$^{\circ}$  \\
KIOST-ESM 	&KIOST/Republic of Korea	&96 × 192&	0.938$^{\circ}$  × 0.938$^{\circ}$  \\
MIROC6 	&MIROC/Japan	&256 × 128	&1.403$^{\circ}$  × 1.403$^{\circ}$  \\
MIROC-ES2L 	&MIROC/Japan	&256 × 128	&1.403$^{\circ}$  × 1.403$^{\circ}$  \\
MPI-ESM1-2-HR 	&MPI-M, DWD, DKRZ/Germany&	384 × 192	& 0.938$^{\circ}$   × 0.939$^{\circ}$  \\
MPI-ESM1-2-LR 	&MPI-M, AWI, DKRZ, DWD/Germany&	192 × 96	&1.9$^{\circ}$  × 1.9$^{\circ}$  \\
MRI-ESM2-0 &	MRI/Japan	&320 × 160	&1.125$^{\circ}$  × 1.125$^{\circ}$  \\
NESM3 	&NUIST/China	&192 × 96	&1.88$^{\circ}$  × 1.88$^{\circ}$  \\
NorESM2-LM &	NCC/Norway	&144 × 96	&2.5$^{\circ}$  × 1.89$^{\circ}$  \\
NorESM2-MM 	&NCC/Norway	&288 × 192&	1.25$^{\circ}$  × 0.94$^{\circ}$  \\
TaiESM1 &	AS-RCEC/Taiwan, China	&288 × 192&	1.25$^{\circ}$  × 0.94$^{\circ}$  \\

    \hline

\end{longtable}

Following data extraction, indices were calculated. The Expert Team on Climate Change Detection Monitoring and Indices (ETCCDI) has led a global effort to develop, compute, and analyze 27 climate indices. This study utilized several extreme precipitation indices as recommended by the ETCCDI. 
\citep{w12030797PrecipitationIndicesTrends}.
These indices include Consecutive Dry Days (CDD), Consecutive Wet Days (CWD), the Number of Heavy Precipitation Days, Wet Day Precipitation, Maximum 1-Day Precipitation, Maximum Consecutive 5-Day Precipitation, and the number of wet and dry days. 
CDD, for example, are used to indicate drought conditions, defined as successive days with less than 1mm of rainfall.
Table \ref{tab:ExtremeIndices} presents a description of each index used within this research.

\begin{longtable}{|p{1.5cm}|p{3.8cm}|p{3.6cm}|p{1.4cm}|}

    \caption{Summary of extreme precipitation indices used in this study 
    \citep{AZMAT2018961}}
    \label{tab:ExtremeIndices} \\

	\hline		
	Index	& Descriptive Name	&Definition &	Units \\
 \hline
CDD &	Consecutive dry Days &	Highest count of successive dry days	& days \\
	CWD	& Consecutive Wet Days
 &	Highest count of successive wet days &	days \\
 WD & Wet Days & number of precipitations days & days\\
	R10 mm	& Number of heavy precipitation days &	Yearly count of days when RR 
 $\ge$ 10 mm &	days \\
PRCPT
& Wet day precipitation &	Annual total precipitation from wet days &	mm \\
	Rx1day	& Maximum 1-day precipitation &	Monthly highest single-day precipitation	& mm \\
	Rx5day	& Maximum consecutive 5-day precipitation	& Monthly highest cumulative precipitation over five days	& mm \\

    \hline


\end{longtable}

The initial step was resampling within the Python framework. 
The historical and Future scenario daily precipitation dataset from GCMs, spanning 138 grid stations for the timeframe 1950–2015, was selected for the complete year without the hydrological seasonal cycle analysis. 

Subsequently, the analysis is conducted using Principal Component Analysis (PCA).
PCA aims to reduce the dimensionality of an extensive dataset consisting of a time series of daily data, 
PCA was employed to transform 
large matrix (CMIP6 Data) into a more compact form while retaining the essential characteristics of the data. 
Data is transformed on an impertinent orthogonal axis known as Principal Components (PC). The PCs were determined by constructing a symmetric covariance matrix and employing a one-dimensional transformation. Eigenvectors indicate the direction, while eigenvalues represent the magnitude of the PC axis, reflecting the variability of the data.
Principal components with a high amount of information, explaining higher cumulative variance, 
were determined based on a scree plot, and these selected components were then utilized for further analysis. PC scores 
represent climate change patterns at specific sites and serve as an alternative to meteorological parameters, offering statistical independence 
\citep{rasmus}. 

Finally, the analysis proceeds with the Agglomerative Herarichal Clustering (AHC) stage. During this phase, 
Clusters exhibiting similar climate signals were successfully identified. Climate change indications, previously estimated using PCA, were represented in the form of component scores. Component scores of the top PCs were then utilized in the Agglomerative Hierarchical Clustering, 
which operates iteratively. This algorithm employs a bottom-up approach, starting from individual points or self-clusters and progressively expanding cluster sizes through the inclusion of nearest points one by one. As a result, various successive combinations of the data point clusters were generated.

An optimal number of clustering hyperparameters was determined based on the Euclidean distance between clusters. 
The dendrogram tree visually explains clustering groups and Euclidean distances among them, forming the foundation for determining an optimal number of clusters. Cluster validity indices assisted in identifying the number of optimum clustering groups.  Corresponding to the optimum clusters for all seasons within a year, the maximum Euclidean distance guided the truncation of the dendrogram. The dendrogram was acquired by AHC. Following that, the count and specific stations constituting each cluster were determined. 



Distinct station clusters were visually depicted on the map of the study area using ESRI ArcGIS. To improve clarity, boundaries representing various climate regions were clearly outlined. A reference station was picked for each climate zone by analyzing the average climatic signals from all grid stations in that zone. The climate of the chosen reference station was considered a stand-in representation of the overall climate within the respective zone.


The silhouette score was computed as the average of the Euclidean distances between clusters. The optimal number of clusters was identified by selecting the one with the highest silhouette score. The silhouette score was calculated using the following formula:
\begin{equation}
    S = \frac{1}{NC} \Sigma \frac{1}{ni} \Sigma \frac{b(r)-a(r) }{max[ b(r),a(r)]} 
\end{equation}


Utilizing Social Economic Pathways (SSPs), the IPCC depicted five climate-forcing scenarios in the Sixth Assessment Report \citep{nuimeprn17886IPCC6ClimateChangeReport2023}.
Two SSP scenarios, SSP245 and 585, were selected out of these, which correspond to mitigation and high emissions scenarios, respectively. These two SSPs were selected to encompass a broad spectrum of greenhouse gas emissions projected in these scenarios and to work on extreme scenarios, which are the best and the worst. 

The daily precipitation data (2015–2099) from selected General Circulation Models (GCM) specific to each zone were extracted for the SSP245 and SSP585 scenarios. This data was then integrated 
to capture anticipated climate trends. The GCM selections from a pool of 23 CMIP6 GCM were based on analysis of precipitation data on a daily basis during both the past phase (1950–2015) and the projected phase (2015–2099) at reference stations. 


Precipitation data was gathered for the distinct future scenarios, namely SSP245 and SSP 585, during the projected phase. The selection of GCM was conducted independently for each future scenario, resulting in two sets of chosen GCMs for each climate zone. This approach aimed to encompass the full range and diversity of climate signals, 
laying the foundation for the envelope-based selection approach.

Following compilation of data from 21 GCMs for both past (1950–2014) and future phase (2015–2098) under SSP245 and SSP585, 
Principal Component Analysis was conducted across all reference stations.
The objective was to reduce the dimensionality of an extensive time series matrix spanning 148 years for each reference station in every season. PCA facilitated the transformation of this large matrix into a more compact form while preserving essential data characteristics.
PCs were derived from the data of 20 GCMs at the corresponding reference stations. The scree plot was utilized to identify highly ranked PCs explaining the maximum cumulative variance in the dataset. 
These component scores provide an alternative representation of climate signals, offering statistical independence from meteorological parameters. 

The variability given by each individual Principal Component ranges between 4\% and 6\%, indicating equal importance in determining the hierarchy of GCM clusters for zone 4. 
The climate signals, represented by component scores from Principal Components (PCs) explaining 95\% of variability within the data, 
were utilized in 
clustering of General Circulation Models. AHC facilitated the formation of clusters with similar descriptive statistics among GCMs. This step enabled the identification of distinct clusters of GCMs exhibiting similar climate characteristics, as detailed in Section 3.1.2. 





The Coupled Models Intercomparison Project (CMIP) has released six generations of datasets. Each generation differs in its methodology for producing this dataset. This research also analyzes the variations in data from the latest generation of CMIP6 compared to the earlier generation of CMIP5 over the Jhelum and Chenab River Basin. In the results section, the outcomes are shown in the form of spatial maps. This step reveals the degree of vulnerability within the region to climate change. 

For comparison, mean operation was carried over model data for each scenario in order to have a generalized value for each type of 
GCM which is a representative of the projected data.
The data was formatted such that in the table, there existed one row for each station and one column for each GCM. Hence there existed 138 rows and 20 columns in the table for each scenario. 
A one-to-one join was performed in ArcGIS Pro after it was validated that each station matched its corresponding station name in the attribute table exactly. 
For spatial interpolation, the Inverse Distance Weighted Averaging (IDW) Tool was employed using its default settings. The parameter values included: power set to 2, standard neighborhood configuration, a maximum of 15 neighbors, a minimum of 10 neighbors, sector type set to 1, and an angle of 0. No weight field was applied, and the output extent was confined to the Basin.
Then using a raster calculator, the difference between the rasters (SSP585 and SSP245), (SSP245 and RCP4.5) was obtained. The method has been used in previous studies for change detection \citep{RasterDifferentation2JAMES2012181}. The differentiation of rasters is useful in finding extreme values as well \citep{RasterSubtractiondoi:10.1080/13658816.2024.2301727}. 
The difference between SSP 585 and SSP 245 presented the difference in projected precipitation for the basin under climate change scenarios.
The contrast between SSP245 and RCP4.5 illustrates the variability in climate forecasts resulting from the use of different generations of CMIP models. This represents the initial regional comparison of newer-generation CMIP models.
In the previous researchers overall comparison of CMIP generations were examined on a national or global scale \citep{BAGCACI2021105576CMIPComparisonTurkey, zhang_uncertainty_2021}.

\section{Results \& Discussion}
\label{sec:results}

This part outlines comprehensive details of the findings acquired by this study. Its goal is to offer a detailed examination and interpretation of the data, shedding light on the key outcomes and their implications. 
Extreme precipitation indices were calculated for CMIP6 GCM which are shown in Table \ref{IndicesTable1}, \ref{IndicesTable2}, and \ref{IndicesTable3}. In these tables, maximum and minimum values are highlighted. These metrics serve as essential tools for studying climate extremes on both global and regional scales \citep{ali_performance_2023}.


\begin{longtable}{|p{3cm}|p{1.3cm}|p{1cm}|p{1cm}|p{1.3cm}|p{1cm}|p{1cm}|}
        \caption{Extreme Indices: Consecutive Wet Days (CWD) and Consecutive Dry Days (CDD)}
    \label{IndicesTable1} \\
    
\hline 
Models& \multicolumn{3}{|c|} {CWD} & \multicolumn{3}{|c|} { CDD}\\
\hline
(Average per year)	& historical	&  ssp245	& ssp585&	 historical &	 ssp245& ssp585\\
\hline
ACCESS\_CM2	&49.1 &	68.7 	&71.7 	&189.3 &	236.0 	&231.0 \\
\textbf{ACCESS\_ESM1\_5}	&\textbf{29.0}& 	78.7 &	\textbf{87.7} &	\textbf{224.9} &	226.7 	&218.2 \\
BCC\_CSM2\_MR	&40.3 &	\textbf{53.5}& 	\textbf{55.5} &	193.4 &	249.9 &	\textbf{249.1}\\ 
CMCC\_CM2\_SR5	&41.5 	&63.6 &	65.0& 	197.7& 	242.3 	&241.6\\ 
\textbf{EC\_Earth3}	&37.4& 	56.1& 	56.3 &	199.8 &	\textbf{250.3} 	&248.4 \\
EC\_Earth3\_Veg\_LR	&41.7& 	57.7 &	58.9 	&195.4& 	245.8 &	244.3\\ 
FGOALS\_g3	&52.4 &	63.1 &	67.9 &	173.1 &	231.1 	&224.1 \\
GFDL\_CM4	&39.7 &	59.4 &	60.3 &	195.6 &	247.8& 	245.4 \\
GISS\_E2\_1\_G	&53.6& 	79.3& 	81.2& 	179.9 &	220.4 &	218.6 \\
INM\_CM4\_8	&52.7 &	\textbf{79.7} &	86.7 &	188.0 &	230.8& 	222.7\\ 
INM\_CM5\_0	&50.5 &	74.0 &	79.4 	&187.6 &	235.7& 	231.2\\ 
IPSL\_CM6A\_LR	&45.5 &	65.0 &	63.2 &	191.2 &	241.4& 	241.8\\ 
KIOST\_ESM	&45.0 &	68.5 &	70.5 &	189.4 &	234.9& 	233.3\\ 
MIROC6	&43.0 &	58.8 &	62.0 &	189.8 &	244.3 	&237.8 \\
MIROC\_ES2L	&\textbf{54.7} &	79.4 &	78.9 &	\textbf{172.9} &	\textbf{217.0}& 	\textbf{217.2} \\
MPI\_ESM1\_2\_LR	&53.8 &	67.9 &	71.4 &	179.9 &	238.6 &	233.7\\ 
MRI\_ESM2\_0	&40.9 &	58.2 	&58.8 &	197.0 &	249.3& 	247.8 \\
NESM3	&53.9 &	72.1 &	77.7 &	181.5 &	234.0 &	228.5 \\
NorESM2\_LM	&51.4 &	76.0 &	75.3 &	190.0 &	233.6&	234.2 \\
TaiESM1	&42.9 &	59.4 &	61.4 &	194.6 	&244.8& 	241.8 \\
\hline

\end{longtable}

\begin{longtable}{|p{3cm}|p{1.3cm}|p{1.2cm}|p{1.2cm}|p{1.3cm}|p{1.2cm}|p{1,2cm}|}
        \caption{Extreme Indices: Wet Days and Total Precipitation}
    \label{IndicesTable2} \\
    
\hline 
Models& \multicolumn{3}{|c|} {Wet Days} & \multicolumn{3}{|c|} { Total Precipitation}\\
\hline
-	&historical	&ssp245&	ssp585&	 historical	& ssp245	& ssp585\\
\hline
ACCESS\_CM2	&141 	&149 	&154 	&721.88 &	853.38 	&892.70 \\
\textbf{ACCESS\_ESM1\_5}	&140 &	167 	&174 &	\textbf{347.04} &	915.97 &	\textbf{1083.67}\\ 
BCC\_CSM2\_MR	&122 	&124 	&126 &	729.67 	&752.06& 	779.68\\ 
CMCC\_CM2\_SR5	&122 	&138 &	139 &	719.89 &	855.51& 	873.80 \\
\textbf{EC\_Earth3}	&\textbf{107} &	\textbf{118} &	\textbf{119} 	&\textbf{730.56} &	849.68 &	892.35 \\
EC\_Earth3\_Veg\_LR	&124 	&132 	&135 	&717.75& 	793.89 &	838.20 \\
FGOALS\_g3	&\textbf{182} &	179 	&\textbf{185} &	681.76 &	\textbf{664.16} &	\textbf{730.62} \\
GFDL\_CM4	&121 &	126 &	128 &	717.41 &	838.79& 	861.64 \\
GISS\_E2\_1\_G	&170 &	183 &	183 &	706.23& 	843.90 &	893.71 \\
INM\_CM4\_8	&155 	&166 &	173 	&705.65 &	\textbf{920.50}& 	1037.51 \\
INM\_CM5\_0	&154 &	162 &	165 &	715.61 &	870.14 	&964.93\\ 
IPSL\_CM6A\_LR	&138 	&142 &	140 	&655.74 &	779.65 &	776.99\\ 
KIOST\_ESM	&144 &	156 	&155 &	725.96& 	876.90 	&917.71 \\
MIROC6	&134 &	138 &	143 &	711.32 &	772.88 &	845.04 \\
MIROC\_ES2L	&179 	&\textbf{186} &	184 	&709.21 &	917.56 &	948.65 \\
MPI\_ESM1\_2\_LR	&162 &	154 	&157 &	717.87 &	732.96 	&793.37 \\
MRI\_ESM2\_0&	122 &	129 	&129& 	706.07 &	789.40& 	797.60 \\
NESM3	&159 &	157 	&160 &	730.19& 	827.65 	&928.37 \\
NorESM2\_LM	&155 	&166 &	164 	&703.57 &	872.27 	&901.79 \\
TaiESM1&	129 &	137 &	139 &	716.04 &	823.04 &	843.06 \\
\hline

\end{longtable}

\begin{longtable}{|p{3cm}|p{1.3cm}|p{1.2cm}|p{1.2cm}|p{1.3cm}|p{1.2cm}|p{1,2cm}|}
        \caption{Extreme Indices: Heavy Rainfall Days and Maximum Precipitation in a day (RX1day)}
    \label{IndicesTable3} \\
    
\hline        
Models& \multicolumn{3}{|c|} {Heavy Rainfall Days} & \multicolumn{3}{|c|} { Max 1-day precipitation 
}\\
\hline
-	& historical& ssp245	& ssp585&historical	& ssp245	&ssp585\\
\hline
ACCESS\_CM2	&18 &	27 &	29 	&1.11 &	1.27 &	1.34 \\
\textbf{ACCESS\_ESM1\_5}	&\textbf{7} 	&29 	&\textbf{35} 	&\textbf{0.87} &	1.26 &	1.29 \\
BCC\_CSM2\_MR	&19 &	25 	&26 &	1.38 &	1.46 &	1.48 \\
CMCC\_CM2\_SR5	&19 &	28 	&29 &	1.38 &	1.32 &	1.28 \\
\textbf{EC\_Earth3}	&\textbf{20} 	&29 	&30 &	\textbf{1.51} &	1.49 &	\textbf{1.75} \\
EC\_Earth3\_Veg\_LR&	18 &	26 	&27 	&1.44 &	1.45 	&1.63 \\
FGOALS\_g3&	15 &	18 	&\textbf{21} 	&1.08 &	1.43 &	1.39 \\
GFDL\_CM4	&19 &	29 &	29 &	1.49 &	1.49 &	1.49 \\
GISS\_E2\_1\_G	&16 	&26 &	28 &	1.18 &	1.30 &	1.24 \\
INM\_CM4\_8	&17 	&\textbf{30} 	&34 &	1.16 &	1.22 &	1.25 \\
INM\_CM5\_0	&17 	&28 &	31 &	1.25 &	1.35 	&1.38 \\
IPSL\_CM6A\_LR	&16 	&25 	&24 &	1.19 &	1.18 	&\textbf{1.05} \\
KIOST\_ESM	&18 &	28 	&29 &	1.44 &	1.60 &	1.59 \\
MIROC6	&18 	&25 	&28 	&1.40 &	1.29 &	1.15 \\
MIROC\_ES2L&	16 	&28 &	29 &	1.08 &	1.42 &	1.42 \\
MPI\_ESM1\_2\_LR	&16 	&\textbf{22} 	&24 &	1.19 &	\textbf{1.04} &	1.17 \\
MRI\_ESM2\_0	&18 &	26 &	27 &	1.47 &	\textbf{1.61} &	1.42 \\
NESM3	&17 	&26 	&30 &	1.17 &	1.21 &	1.15 \\
NorESM2\_LM	&17 	&27 	&28 	&1.38 &	1.34 &	1.34 \\
TaiESM1	&19 &	27 &	28 &	1.28 &	1.49 &	1.49 \\
\hline

\end{longtable}

The models reveal a growing trend in the frequency of Consecutive Wet Days and Consecutive Dry Days, as illustrated in Table \ref{IndicesTable1}. Additionally, Table \ref{IndicesTable2} highlights an uptick in the occurrence of wet days and periods of heavy rainfall. Similarly in table \ref{IndicesTable3}, rise in the number of days with intense rainfall. 
This rise in the magnitude of extreme indices indicates that, due to climate change, more severe precipitation events will occur in the future. 
Through Extreme Indices, it was known that the GCMs produced the 
extreme values are ACCESS\_ESM1\_5 and ECEarth3. The results from the extreme indices through CMIP6 GCMs support the conclusions drawn by \citet{HUSSAIN2023106873PrecipitationNDVI}. Conversely, the findings contrast to the Upper Indus River Basin, which is a neighboring river basin where extreme precipitation indices chosen to analyze the spatiotemporal distribution of precipitation extremes in the region show inconsistent and heterogeneous patterns \citep{w12123373QamarUpperIndus} as most of GCM's show consistent pattern in Jhelum and Chenab River Basin. This difference could be because of the difference in land cover in the two river basins. These results aid in assessing CMIP6 GCMs to produce precipitation indices across the complex regions of the Jhelum and Chenab river basins. It can provide a basis for future studies on extreme weather events in the region. For future research, these precipitation indices can be used for trend detection \citep{hussain_regional_2013Trends, ExtremePrecipitationIndicesatmos14020210}. 
This theorem of trend detection through extreme indices was presented by \citet{Trenddoi:10.1080/01621459.2019.1705307}.

For the envelope-based selection of GCMs, regionalization was done in order to have a more precise selection \citep{atmos14101497}.  The regionalization process involves two key steps: Principal Component Analysis (PCA) and Agglomerative Hierarchical Clustering. The patterns of climate change were illustrated through PCA applied to daily precipitation data from 138 sites, covering the period from 1950 to 2015. 
Subsequently, Agglomerative hierarchical clustering was employed to identify clusters of sites manifesting similar climates and characteristics. The results of every stage in the AHC process are outlined below.

After the execution, 23 significant principal components were identified, collectively explaining 95\% of the cumulative variance 
over the study area. 
The subsequent sections detail the cumulative variance explained by twenty principal components. According to these findings, the initial 20 principal components accounted for approximately 94–95\% of the variance in each season. Subsequently, these 20 principal components were utilized in the regionalization process for agglomerative clustering analysis. 
For each station, component scores were computed for every principal component, representing the climate signals produced at each corresponding location.
Agglomerative Hierarchical Clustering was employed to cluster the component scores of sites 
based on the first 20 PCs. 
Determining the optimum cluster 
was facilitated by silhouette score, a metric to check the validity of clusters,
where the highest score corresponds to the optimal clustering. The test results recommended optimal clustering into 10 groups.
The maximum Euclidean distances associated with the optimal number of clusters were found to be 95.
Dendrogram trees generated through Agglomerative Hierarchical Clustering (AHC) were then obtained. 
The dendrogram trees produced by Agglomerative Hierarchical Clustering were cut at a Euclidean distance of 95 to determine the optimal clusters and the station locations within each one.


On the geographical map, stations belonging to each cluster were marked, with each cluster represented by a distinct circular symbol. Each reference station had characteristics for the cluster indicating its homogeneous climate. To enhance clarity, cluster boundaries were delineated, effectively dividing the region into distinct climate zones. 
After incorporating outliers into the closest clusters, the river basins were divided into 10 groups.    
Stations within each cluster, representing an approximate average of climate signals from the included stations, were designated as reference stations. 
Reference stations were designated for each group or climate zone.

\begin{figure}[H]
    \centering
    \includegraphics[width=1\linewidth]{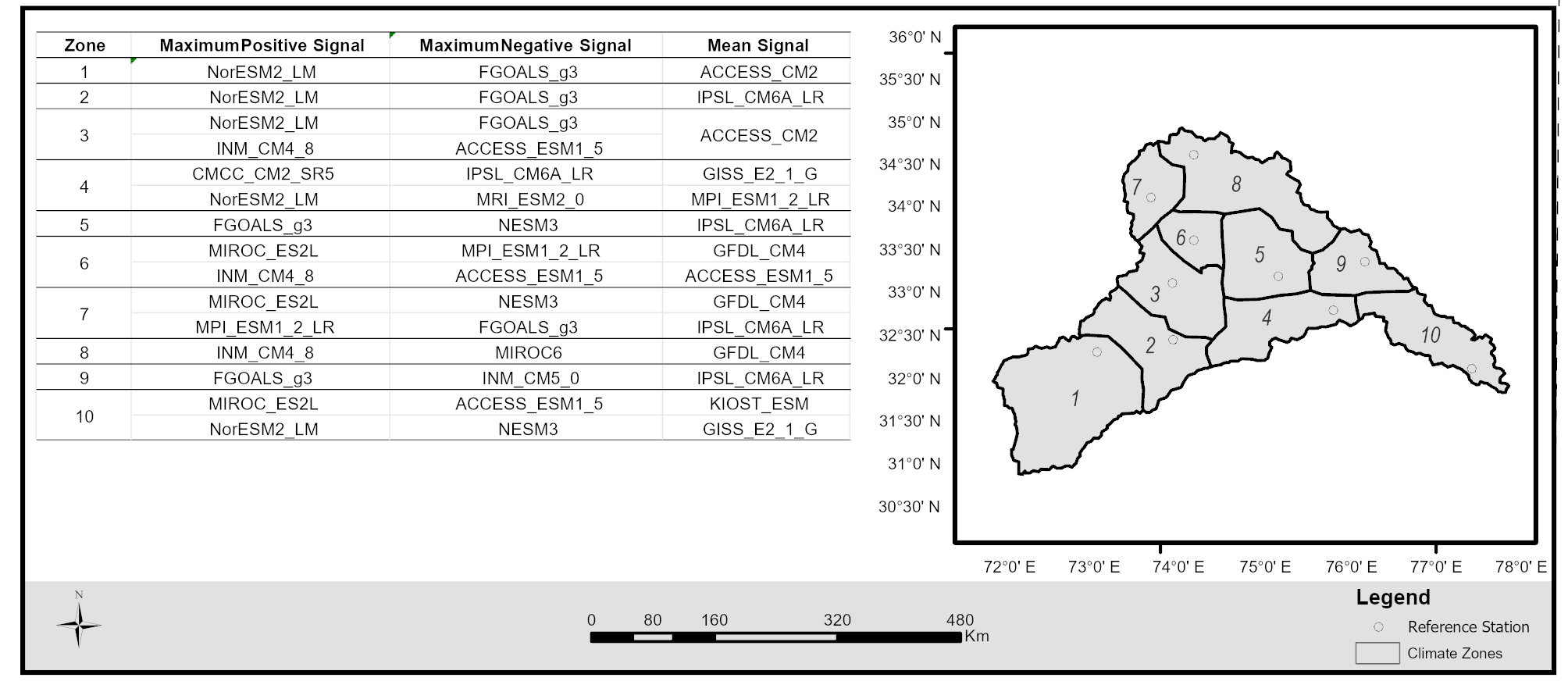}
    \caption{Selected CMIP6 GCM's for each climate zone}
    \label{fig:enter-labelSelectedGCMs}
\end{figure}

GCMs were chosen for each climate zone and season within the study area using the envelope-based approach. The selected GCMs, utilizing the past (historical) and future phase data 
(SSP245 and SSP585) from 20 GCMs across various climate zones and seasons, are listed in figure \ref{fig:enter-labelSelectedGCMs}. 

Climate signal was calculated for each GCM after AHC.
Some of the zones are listed with more than one selected GCM because these GCMs had very close values for extreme signals. Therefore, any of the two selected GCMs can be used as representative GCMs for the corresponding zone as shown in the map in Figure \ref{fig:enter-labelSelectedGCMs}.
For the selection of GCM for the complete basin,  
was obtained for the performance of each GCM. Selected GCM for the highest positive signal is NorESM2\_LM, while for the highest negative signal is FGOALS\_g3, while the mean signal near zero is given by IPSL\_CM6A\_LR. These two GCMs present the NorESM2\_LM and FGOALS\_g3 precipitation fluctuations for the Jhelum and Chenab River Basins. This selection of GCM is made through an envelop-based approach.

\begin{figure}
    \centering
    \includegraphics[width=1\textwidth]{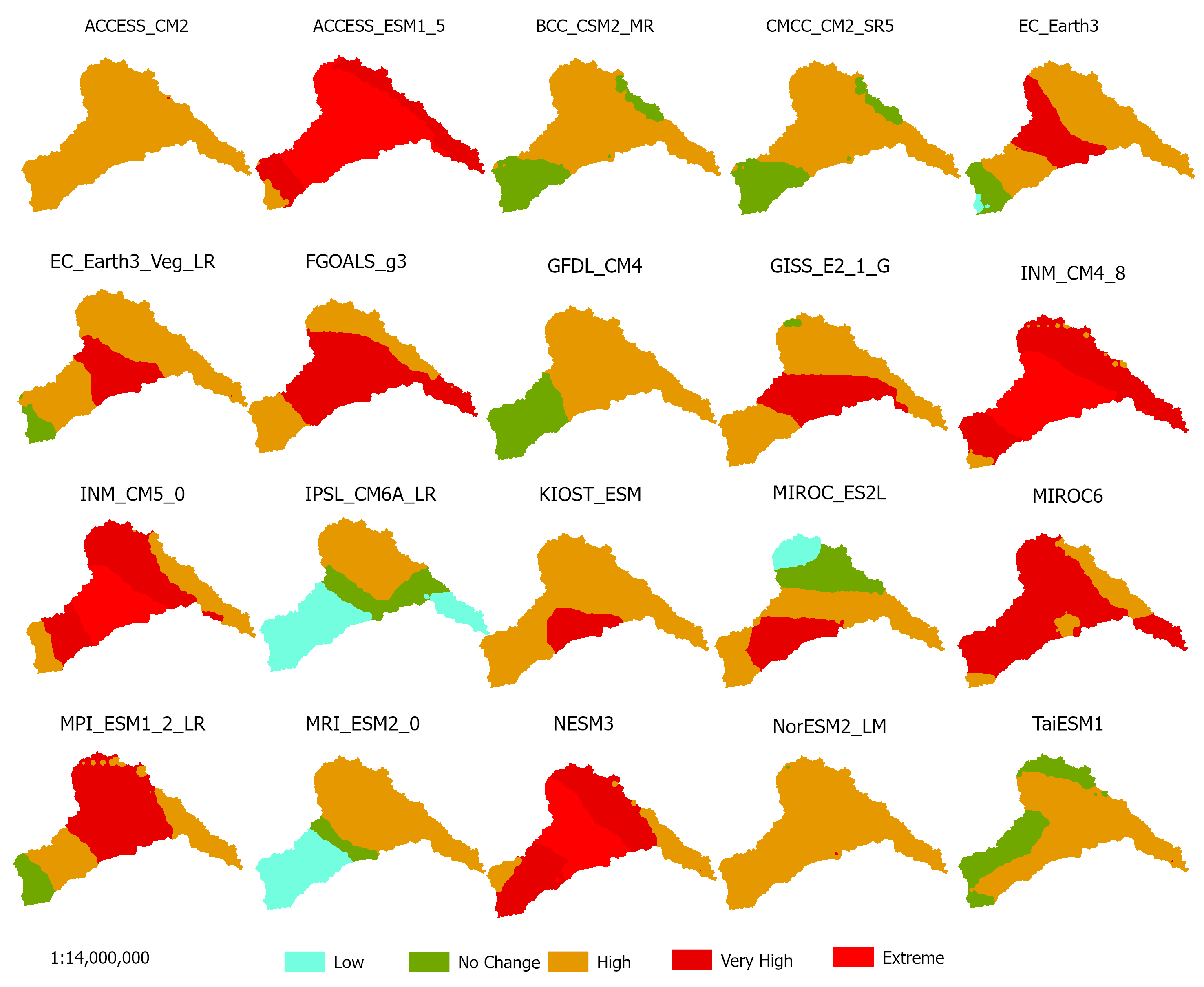}
    \caption{SSP's Variability Map showing the effect of climate change under SSP scenarios }
    \label{fig:ClimateChangePic1}
\end{figure}

For comparison of CMIP6 SSP245 and SSP585, CMIP6 data was formatted such to be able to load it with point shapefile in ArcGIS Pro software. 138 sampling points in shapefile format as used in previous research for CMIP5 data by \citet{atmos13020190}, were used in this research as well. 
A mean operation was performed on the GCM data over the 83 years.
This comparison was particularly made to indicate the areas that would be affected the most due to climate change. Based on CMIP6 data, the variation between the most optimistic and the most pessimistic scenarios for future projections indicates the changes in precipitation due to climate change. Figure \ref{fig:ClimateChangePic1} presents an SSP Differential Map depicting the variations in precipitation across CMIP6 climate change scenarios. In figure \ref{fig:ClimateChangePic1}, shades of the regions show the extent of precipitation difference under CMIP6 GCM's climate change scenarios. 



In figure \ref{fig:ClimateChangePic1}, the blue part shows the area where average precipitation is less than -1, the green parts where the average difference is between -1 and 1, the orange parts show the average difference between 1 to 5, brown part shows the average difference between 5 to 10, and red part shows the area where average difference is greater than 10. This difference was observed from -5 up to the maximum value of 24. The majority of GCMs forecast a rise in precipitation.
ACCESS\_ESM1\_5 
provides the extreme difference between SSP585 and SSP245.

It was inferred from figure \ref{fig:ClimateChangePic1}, that the Jehlum and Chenab river basin would be severely effected with Climate Change. 
The red-shaded area will be most affected due to the increase in precipitation. It depicts extreme precipitation differences due to extreme heating. Most of the GCMs show red areas within the Jhelum and Chenab basins where the average precipitation difference is greater than 10 millimeters. This indicates a large difference in mean precipitation due to climate change. The red regions are mostly at higher altitudes with mountain ranges and glaciers. The area includes most parts of Jammu and Kashmir and some parts of Punjab province. Srinagar, Muzaffarabad, and Wazirabad are large cities located in this area. 
More frequent flash floods are expected in these areas in the future.

Global warming is causing disruption in normal ecosystems. The increase in precipitation is a result of an increase in temperature. Hence, the regions depicted in red are the regions that will have a sharp increase in precipitation within the coming years in case the phenomena of climate change get worse and no action is taken on emissions reduction and climate change. 
An increase in temperature would also result in more evapotranspiration and increased snowmelt in high-altitude regions. This can result in higher stream flows within the river for a certain span of years till all the glaciers get melted. For these years of increased river flow, huge floods can be expected. 
When the Himalayan region is left with no glaciers, in summer, the streams could get dry in case of no precipitation. Hence, causing ecosystem disruption, leading to droughts and reduced agriculture within the region.

\begin{figure}[H]
    \centering
    \includegraphics[width=1\textwidth]{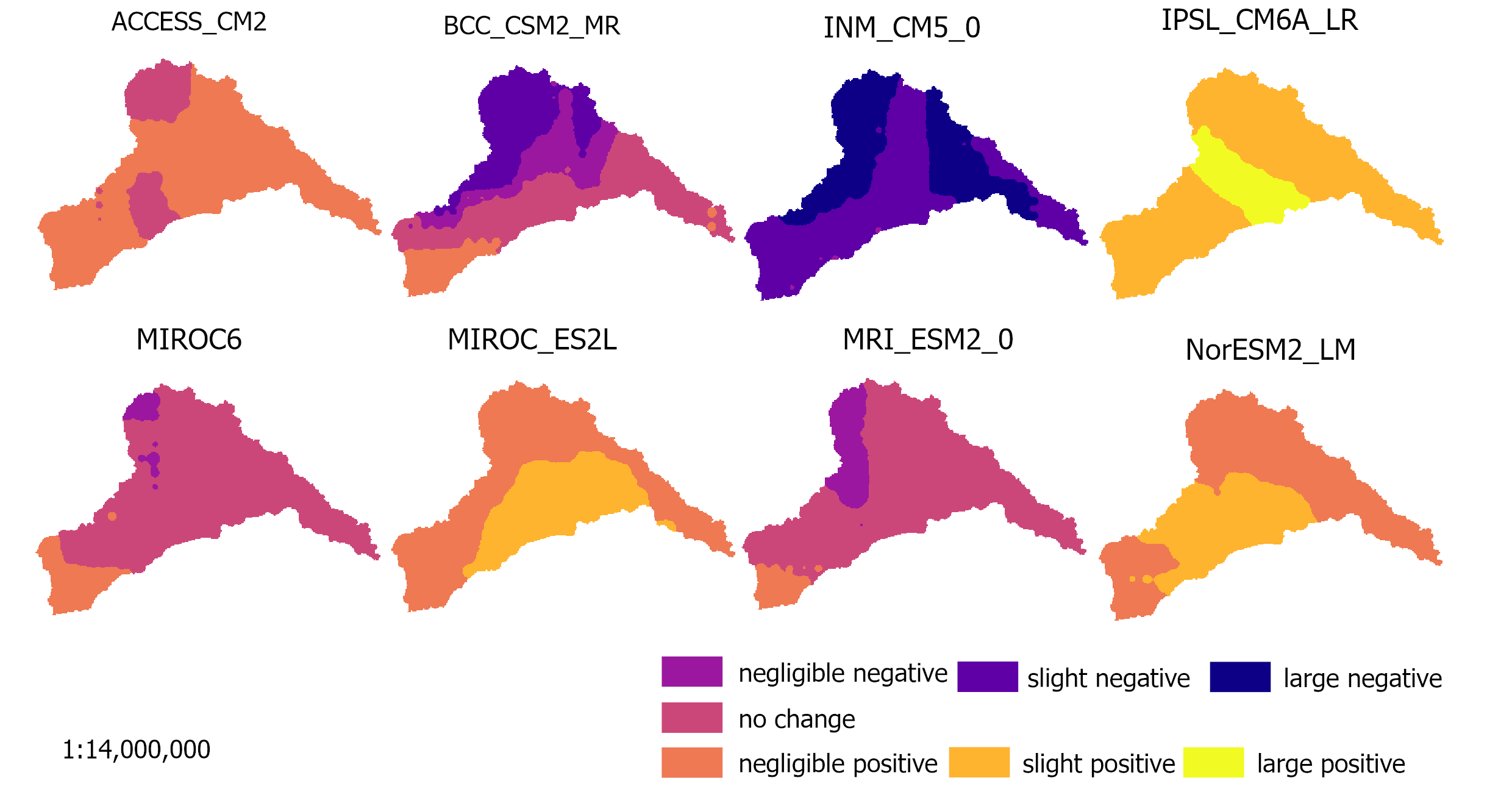}
    \caption{Variability of CMIP5 and CMIP6 Data in Jhelum and Chenab River Basin }
    \label{fig:enter-labelDifferenceCMIP56}
\end{figure}

For the difference check between CMIP5 and CMIP6, the GCM, which were common in the two by having the following product, were scrutinized. A total of 7 GCMs were found. The same procedure for pre-processing was adopted for CMIP5 as well.
During filtering of the same interval for the two datasets, it was known that even within the 2015 to 2098 interval, CMIP5 and CMIP6 had a difference of 21 days.  So the number of days from 2015 to 2098 for CMIP5 was 30659 while for CMIP6 was 30680. This difference was initially thought to be because of the consideration of leap-year dates. However, it was disclosed that CMIP5 data had missing values for 27 February of every leap year. 
Similar inconsistencies were also found in CMIP6 data. These places were filled by interpolation.
Figure \ref{fig:enter-labelDifferenceCMIP56} presents the difference between the CMIP6 SSP scenario and the corresponding CMIP5 RCP scenario.


With the comparison of CMIP6 and CMIP5 data, 
no significant difference in mean precipitation was observed for most parts of the region for all GCMs due to a small difference in mean precipitation within the region. Hence, previous research conducted using CMIP5 data stands valid. Using CMIP6 data in previous research in most cases would produce in similar results.
INM\_CM5\_0 and IPSL\_CM6A\_LR, however, show significant negative and positive differences, respectively, for a small part of the region. 
After the comparison of SSP245 and RCP4.5, similar conclusions were drawn through the comparison of SSP585 and RCP8.5 as well.



\section{Conclusion}
\label{sec:conclusion}



Since the state-of-the-art CMIP6 models were made publicly available, there was a need to know which of them are best suitable for the region. 
The current world 
lacks a consensus to have criteria 
for selecting GCMs that are accepted worldwide as a permanent reference.  
This research explored statistical and dynamic approaches to calculate 
variability in climate change forecasts. 
In an effort to improve reliability with climate models 
across diverse climate 
zones, 
the study focuses on calculating extreme indices for the latest CMIP6 models. These indices are necessary for the appropriate planning and management of regional water resources.  
Through Extreme Indices, it was inferred that Australian ACCESS\_ESM1\_5 and European Consortium ECEarth3 are the GCM’s
producing extreme values.

Envelope based technique 
was used 
for the final selection of the General Circulation Model.
The selection relies on daily seasonal precipitation 
derived from 
coupled models gridded data at 138 sampling points within the Jhelum and Chenab River Basin. The data used covers the historical time period from 1950 to 2015 for the two main rivers of Punjab province. 
The data extracted for the baseline period spanned from 1950 to 2015, while the data for the two future scenarios covered the years 2015 to 2099. The Principal Components encapsulated the indicators of climate variations.
The hierarchical agglomerative method applied to the principal components led to the grouping of GCMs based on climate uniformity.
Hence, the resulting zones shared a homogeneous variability in climate signals. 
Using the envelope-based selection criteria, the GCM model exhibiting extreme values within each zone was then selected for that particular climate zone.
The Norwegian NorESM2\_LM and Chinese FGOALS\_g3 and French IPSL\_CM6A\_LR GCMs were selected models for exhibiting the highest positive, highest negative, and mean signal, respectively. These models can be used in future hydroclimate research in the study area to incorporate the maximum precipitation variability of CMIP6 models.

The SSP's Difference Map depicted the area 
under greater threat 
under projected climate change scenarios. 
Through comparison of CMIP5 and CMIP6 GCMs, it was revealed that no significant difference was found between the precipitation projection of RCP and SSP scenarios. Hence, both generations of CMIP data can be used for hydrological analysis within the Jhelum and Chenab River Basin and the release of newer generation CMIP6 data does not out-date the older CMIP5 data. However, in this comparison, mean precipitation values were used for the comparison. For future work, through further statistical methods, a more detailed comparison can be drawn to map the difference between the CMIP data for RCP and SSP Scenarios.

.

\textbf{Data Availability:} The dataset can be accessed on the NASA Center for Climate Simulation website \url{https://www.nccs.nasa.gov/services/data-collections/land-based-products/nex-gddp-cmip6}. The scripts and code developed during this research is accessible via the GitHub repository \url{https://github.com/SaadAhmedJamal/ClimateModel_Selection_Envelop_Approach}.

\backmatter

\bibliography{ref.bib}

\end{document}